\documentclass{INTERSPEECH2023}

\usepackage{booktabs}

\usepackage{cleveref}

\interspeechcameraready

\title{Focus the Sound around You: Monaural Target Speaker Extraction via Distance and Speaker Information}

\name{Jiuxin Lin$^{1, *}$, Peng Wang$^{2, *}$\thanks{$^{*}$ Equal contribution.}, Heinrich Dinkel$^2$, Jun Chen$^1$, Zhiyong Wu$^{1, \dagger}$\thanks{$^{\dagger}$ Corresponding author.}, \\Yongqing Wang$^2$, Zhiyong Yan$^2$, Junbo Zhang$^2$, Yujun Wang$^2$}

\address{
  $^1$Shenzhen International Graduate School, Tsinghua University, Shenzhen, China\\
  $^2$Xiaomi Inc., Beijing, China}
\email{linjx21@mails.tsinghua.edu.cn, zywu@sz.tsinghua.edu.cn}
\begin{document}
    
\maketitle

 \def\model{NS-Extractor}
\begin{abstract}
    Previously, Target Speaker Extraction (TSE) has yielded outstanding performance in certain application scenarios for speech enhancement and source separation. 
    However, obtaining auxiliary speaker-related information is still challenging in noisy environments with significant reverberation.
    Inspired by the recently proposed distance-based sound separation, we propose the near sound (NS) extractor, which leverages distance information for TSE to reliably extract speaker information without requiring previous speaker enrolment, called speaker embedding self-enrollment (SESE). 
    Full- \& sub-band modeling is introduced to enhance our \model{}'s  adaptability towards environments with significant reverberation. 
    Experimental results on several cross-datasets demonstrate the effectiveness of our improvements and the excellent performance of our proposed \model{} in different application scenarios.

\end{abstract}

\section{Introduction}
\label{intro}
Target Speaker Extraction (TSE)~\cite{8462661}, also known as Target Speech Extraction, is an essential task in the field of audio processing that involves separating a speech signal of a specific speaker from an audio mixture containing multiple speakers. 
This task has become increasingly important in recent years with the rise of various speech-based applications such as speech recognition~\cite{zhang2018deep}, speaker verification~\cite{michelsanti2017conditional}, and audio conferencing. 
While blind speech separation (BSS) is limited by permutation invariant training (PIT)~\cite{yu2017permutation}, TSE methods face no such restriction.
Moreover, while TSE can extract the desired speaker's speech directly, BSS outputs several speech signals from different speakers, which requires manual selection. 
Nevertheless, TSE has a disadvantage: auxiliary information related to the target speaker such as enrolled voice~\cite{xu2020spex, ge2020spex+, ge2021multi} or lip movements~\cite{pan2021muse,9887809,pan2022selective} are required in advance.
Typically, this necessitates allocating additional resources and encroaching upon the privacy of the information involved.

Recently,~\cite{patterson22_interspeech} proposed distance-based sound separation (DSS), which can separate monaural audio sources by the perceived distance (due to reverberation) between a listener and a sound emitter. 
DSS produces two audio signals, one from within a fixed threshold distance (``near'') and another from outside the distance (``far''). 
Currently, DSS may face certain limitations in practical applications. 
First, the threshold distance for separation cannot be arbitrarily changed during inference, which might result in having multiple ``near'' sources due to an intrusive sound source coming into the threshold distance range.
As an example, within a meeting, multiple sources might be of equal distance to the microphone, which the approach in~\cite{patterson22_interspeech} is unable to separate.
Furthermore, due to the heavy reliance on the reverberation effect, distance-based separation is limited to smaller rooms with a longer reverberation time (RT60), while many offices are in large rooms with a faint reverberation effect. 
Lastly, previous works based on LSTM~\cite{6795963} can be further optimized to use more modern separation models, which could significantly enhance the user experience. 
Our work is inspired by the human perception of the cocktail party problem, where humans can selectively focus on a specific sound source (i.e., speaker) if it is closer to them, while still filtering noise from far away sources.
Thus we believe that if we incorporate this distance-based source separation into TSE, we can achieve a more potent separation performance.

Although separating mixed audio signals with and without reverberation may appear to be similar tasks, there are significant differences between the two in practice. 
Reverberation can cause several issues in speech modeling~\cite{6296524}, including:
\begin{enumerate*}[label=(\alph*)]
 \item Create echoes that overlap with the original speech signal;
 \item Dampen the high-frequency components of the speech signal;
  \item  Introduce a delay between the original speech signal and the reverberant sound.
\end{enumerate*}
All these may lead to a more difficult understanding of speech. 
Therefore, when conducting TSE in a reverberant environment, a different approach must be taken compared to regular TSE.

While time-domain approaches have seen success on commonly used benchmark datasets such as WSJ0-2mix~\cite{7471631}, some of them such as Conv-TasNet~\cite{luo2019conv} generally perform poorly when faced with reverberant audio~\cite{maciejewski2020whamr}.
This performance decay has been analysed in~\cite{han2022dpccn}, where 
 time-frequency (spectral) domain frameworks have been seen to offer superior separation performance.
Additionally, it was indicated that a sub-band model is capable of modelling the  reverberation effect by focusing on the temporal evolution of the narrow-band spectrum in the results of~\cite{hao2021fullsubnet}.

In this work,   we propose the \textbf{N}ear \textbf{S}ound Extractor (\model{}), a TSE model combing full-, sub-band modeling and speaker embedding self-enrollment (SESE). 
\model{} utilizes the perceived distance to the target speaker as a cue to extract a self-enrolled speaker embedding that represents the voice print of the target speaker, which is then used for further extraction. 
Full- and sub-band modeling are integrated to attain greater stability in extraction performance.
Experimental results show that our proposed \model{} not only outperforms the baseline in terms of signal and perceptual quality but also exhibits superior performance in more complex scenarios. 
\begin{figure*}[htbp]
	\centering
	\subfigure[Distance-based sound separation]{\label{fig1:a}
		\includegraphics[scale=0.45]{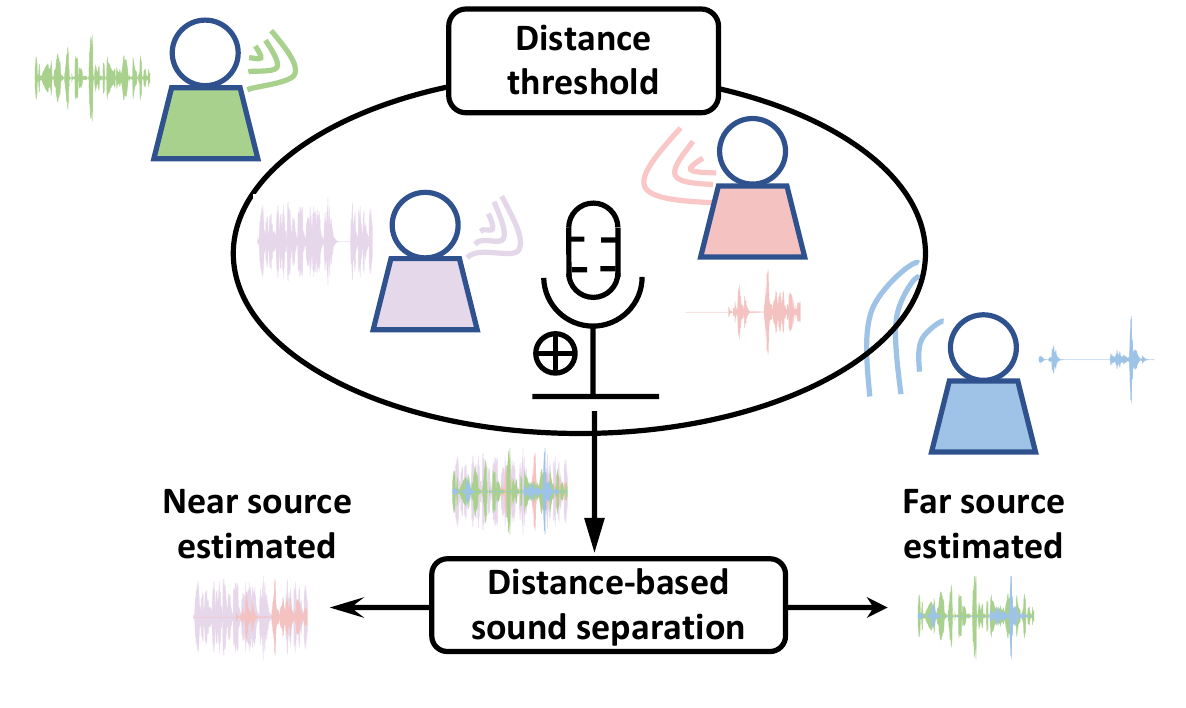}}
	\subfigure[TSE]{\label{fig1:b}
		\includegraphics[scale=0.45]{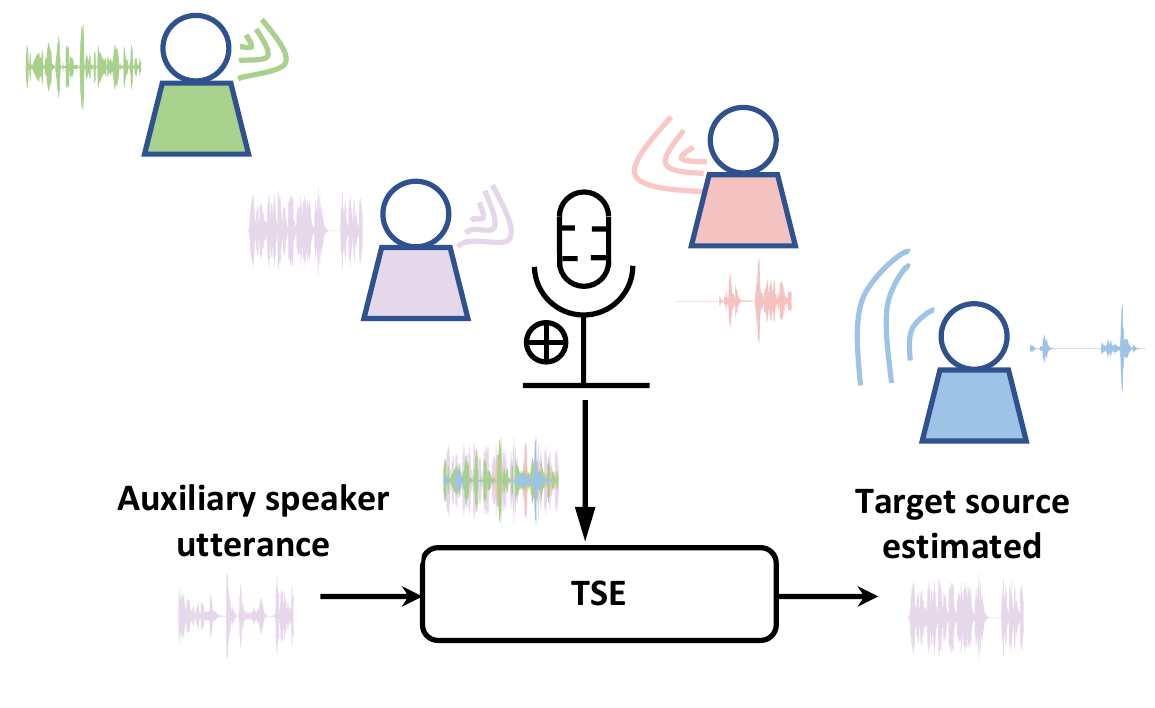}}
	\subfigure[Proposed \model{}]{\label{fig1:c}
		\includegraphics[scale=0.45]{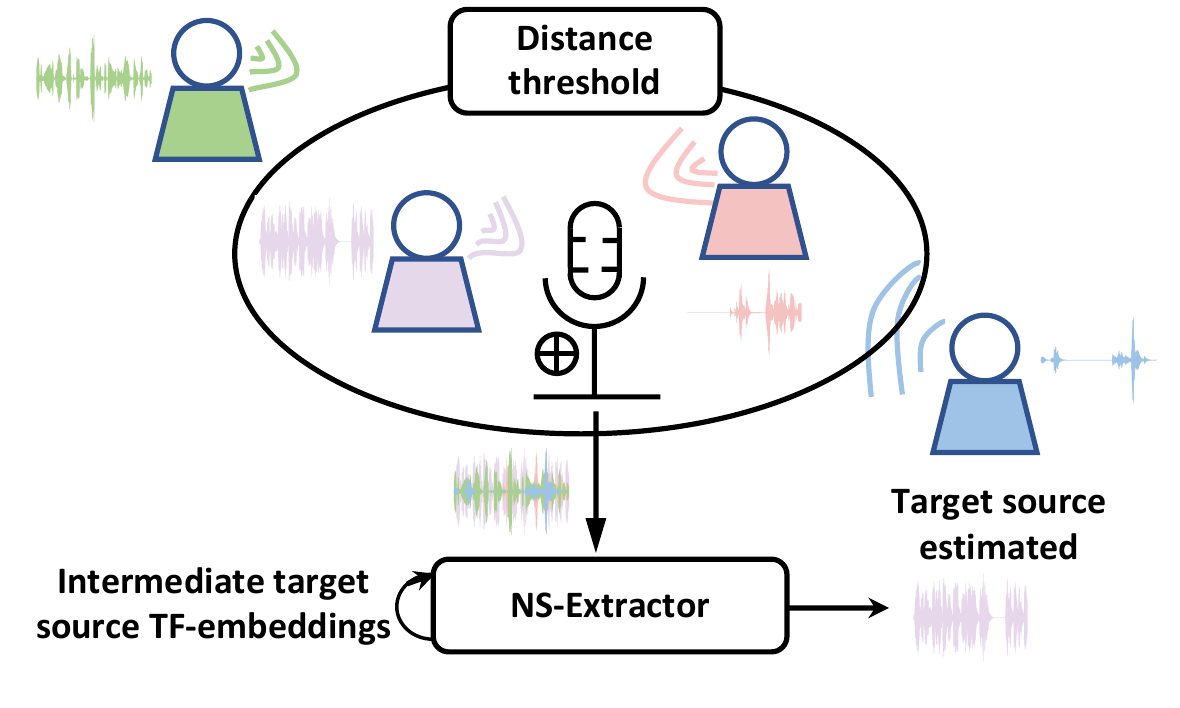}}
        \vspace{-2mm}
	\caption{Illustrations of distance-based sound separation, TSE and our proposed \model{}.}
	\label{fig1} 
 \vspace{-5mm}
\end{figure*}   

\section{Methodology}
\label{sec:methodology}
\subsection{Problem description}
Assuming a $K$-speaker mixture recorded in anechoic conditions, one can formulate the physical model in the time domain as $\boldsymbol{x}[t] = \sum_{k=1}^K \boldsymbol{s}^{(k)}[t]$, where $\boldsymbol{x}$ represents the mixture and $\boldsymbol{s}^{(k)}$ source $k$ in this mixture, and $t$ indexes $T$ time samples. 
The sound envisioned in our work is emitted in a confined space, where each source can be formulated as $\boldsymbol{s}^{(k)}=\boldsymbol{d}^{(k)}\star \boldsymbol{r}^{(k)}$. 
$\boldsymbol{d}^{(k)}$ and $\boldsymbol{r}^{(k)}$ represent the direct-path signal and reverberation, respectively and convolution is denoted by $\star$.

In order to provide a clearer exposition of our work, we provide a comparative analysis between our approach, traditional TSE techniques, and distance-based sound separation, highlighting all their discrepancies. 
Illustrated in~\Cref{fig1:a}, distance-based sound separation in~\cite{patterson22_interspeech} separates mixed audio based on the distance of sound sources in space, which can be expressed as:
\vspace{-2mm}
\begin{equation*}
    \boldsymbol{x} \longrightarrow \sum_{k_i}^{K_{\text{near}}} \boldsymbol{s}^{(k_i)} + \sum_{k_j}^{K_{\text{far}}} \boldsymbol{s}^{(k_j)},
    \vspace{-1.5mm}
\end{equation*}
where the two terms are the sum of near and far targets' sounds respectively. 
This modeling approach also indicates that the estimated targets (near, or far) may contain more than one sound (multiple speakers).
By leveraging the auxiliary speaker-related information provided, TSE~(\Cref{fig1:b}) is capable of extracting the target speech from mixed audio. 
The process can be depicted as follows:
\begin{equation*}
    \boldsymbol{x} \stackrel{\boldsymbol{a}}{\longrightarrow}  \boldsymbol{s}^{(k_g)},
\end{equation*}
where $\boldsymbol{a}$ is the auxiliary speaker-related information, $\boldsymbol{s}^{(k_g)}$ represents the target speech of one single speaker who indexs $k_g$. 
As illustrated in~\Cref{fig1:c}, our proposed \model{} possesses the ability to exclusively extract a single target speech within close proximity using an enrolled speaker embedding, which is obtained from the intermediate target source T-F embeddings.
Thus, additional auxiliary speaker information is not required.
The detailed process will be described in~\Cref{sec:SESE}.

%
\subsection{\model{}}
\label{sec:model}
Our extractor model is based on performing complex spectral mapping~\cite{williamson2015complex, tan2019learning,wang2021multi}, whereby the real and imaginary (RI) components of $\mathbf{X}\in\mathbb{R}^{ 2\times F\times T }$ are concatenated to form the input features, which are then utilized to predict the RI components of each speaker $\mathbf{S^{(c)}}\in\mathbb{R}^{2\times F\times T }$. 
Adhering to the methodology of TF-GridNet~\cite{wang2022tf}, our proposed \model{} first employs 2D Convolution (Conv2D) with a $3 \times 3 $ kernel and global layer normalization (gLN) to compute $D$-dimensional embeddings for each T-F unit $\mathbf{H_x^{(\text{1})}} \in \mathbb{R}^{D\times F\times T}$. 
$\mathbf{H_x^{(\text{1})}}$ is then fed into $C$ stacks of extractor blocks, with each consisting of SESE and full- \& sub-band modeling to refine the T-F embeddings progressively. 
The extractor outputs $\mathbf{\widehat{H_x}}$, a 2D deconvolution (Deconv2D) with 2 output channels and a $3\times 3$ kernel followed by linear activation is then used to obtain the predicted RI components $\mathbf{Y}\in\mathbb{R}^{ 2\times F\times T }$ from $\mathbf{\widehat{H_x}}$.
\subsubsection{Speaker embedding self-enrollment}
\label{sec:SESE}
Each SESE step includes both speaker encoding and speaker embedding fusion.
At each block of the extractor, the input $\mathbf{H_x^{(\text{c})}}$ is chained to the output of the preceding block, while $\mathbf{H_x^{(\text{1})}}$ is directly obtained by encoding the original mixed input spectrum $\mathbf{X}$. 
The input of speaker encoder $\mathbf{R^{\text{(c)}}}\in \mathbb{R}^{F\times T}$ is derived from $\mathbf{H_x^{(\text{c})}}$ through a $1\times 1$ Conv2D.
The speaker encoder consists of a stack of 3 residual blocks followed by an adaptive average pooling layer (AvgPool)~\cite{ge2020spex+}. 
The 1D-AvgPool layer, with a kernel size of 3, compresses the temporal dimension of speaker embeddings in extractor block $c$. 
The resulting single vector $\mathbf{E^{(\text{c})}}\in \mathbb{R}^{1\times F}$, serves as an speaker identity encoding.

Prior to the speaker embedding fusion, a concatenation of speaker embeddings $\mathbf{E^{(\text{c})}}$ and T-F embeddings $\mathbf{H_x^{(\text{c})}}$ is required. $\mathbf{E^{(\text{c})}}$ is replicated across temporal dimension and concatenated with $\mathbf{H_x^{(\text{c})}}$ along dimension $D$ to form a tensor with shape $(D+1) \times T \times F$. 
Conv2D with a $1\times 1$ kernel is employed to restore the dimension to $D \times T \times F$.
\begin{equation*}
   \mathbf{\dot{H}_X^{(\text{c})}} = \operatorname{Conv2D}(\operatorname{Concat}(\mathbf{H_x^{(\text{c})}}, \mathbf{E^{(\text{c})}}), D+1, D) \in \mathbb{R}^{D \times T \times F},
\end{equation*}
where $D+1$ and $D$ represent the number of input and output channels respectively.

The speaker embedding fusion block is employed to model the internal relationship inside $\mathbf{\dot{H}_X^{(\text{c})}}$. 
The input tensor $\mathbf{\dot{H}_X^{(\text{c})}}\in \mathbb{R}^{D \times T \times F}$ is viewed as $T$ separate sequences, each with length $F$.
To model, the local relationship between a speaker and spectral information at the frame level, a single-layer bidirectional LSTM (BLSTM) architecture is utilized. 
The unfold and layer normalization (LN) operation in~\cite{wang2022tf} are employed as follows: 
\vspace{-0.5mm}
\begin{scriptsize}
\begin{equation*}
\begin{split}
& \mathbf{U^{(\text{c})}}=\left[\operatorname{Unfold}(\mathbf{\dot{H}_x^{(\text{c})}}[:, t,:]), \text {for } t=1, \ldots, T\right] \in \mathbb{R}^{(I \times D) \times T \times \frac{F}{J}}, \\
& \mathbf{\dot{U}^{(\text{c})}}=[\operatorname{BLSTM}(\operatorname{LN}(\mathbf{U^{(\text{c})}})[:, t,:]),\text{ for }t=1, \ldots, T] \in \mathbb{R}^{2 H \times T \times\frac{F}{J}},
\end{split}
\end{equation*}
\end{scriptsize}
\begin{figure}[t]
\centering
\includegraphics[width=0.6\linewidth]{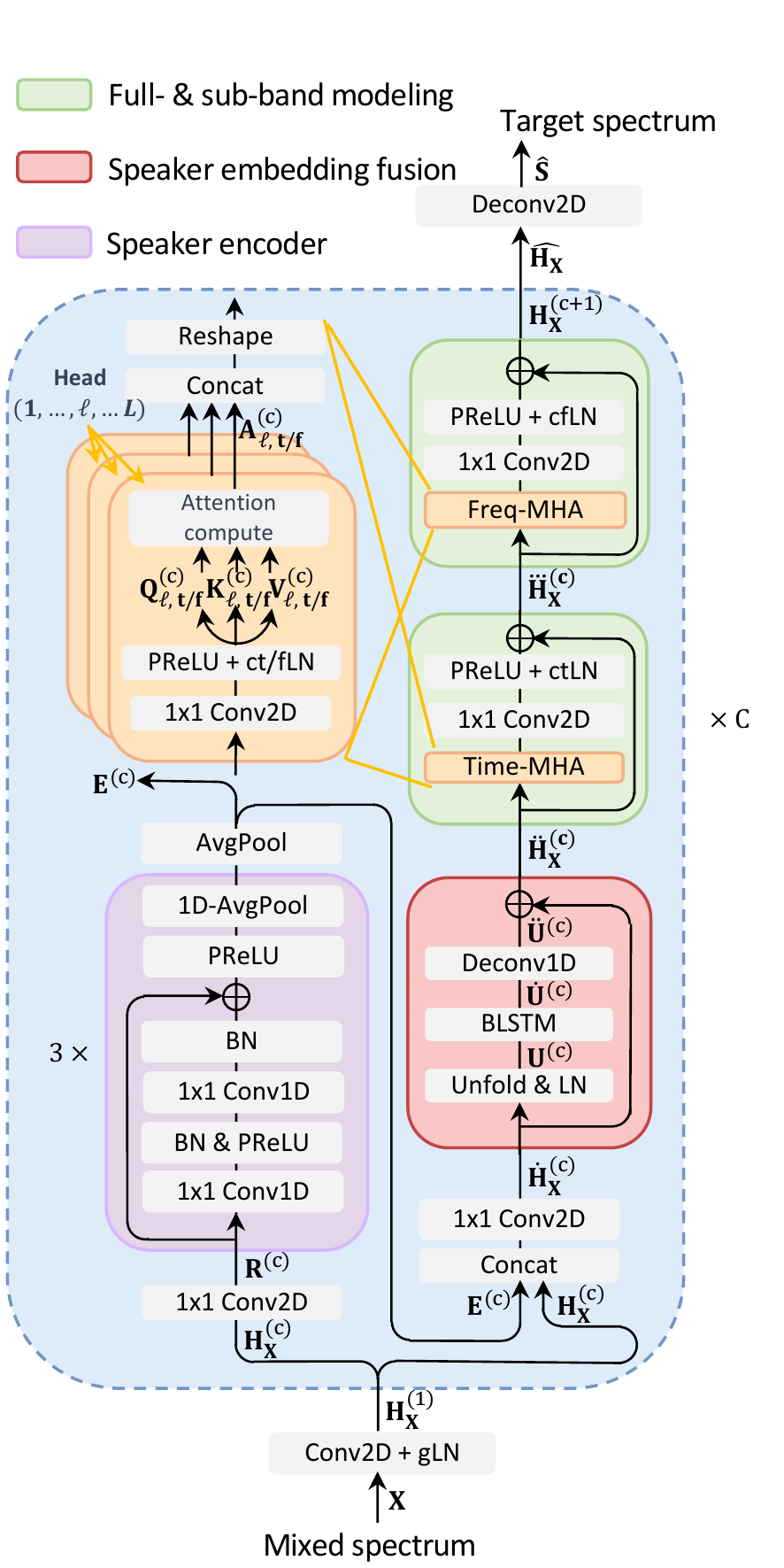}
\caption{Detailed structure of proposed \model{}. The whole extraction process consists of three steps: self-enroll speaker encoder, speaker embedding fusion and full- \& sub-band modeling.}
\label{fig2}
\vspace{-2em}
\end{figure}
where $I$ and $J$ represent kernel size and stride size respectively, $H$ denotes the number of hidden units in BLSTMs in each direction.
Subsequently, a 1D deconvolution (Deconv1D) layer with kernel size $I$, stride size $J$, input channel $2H$ and output channel $D$ is applied to the hidden embeddings of the BLSTM:
\begin{footnotesize}
\begin{equation*}
\nonumber
\mathbf{\ddot{U}^{(\text{c})}}=[\operatorname{Deconv} 1 \mathrm{D}(\mathbf{\dot{U}^{(\text{c})}}[:, t,:]), \text { for } t=1, \ldots, T] \in \mathbb{R}^{D \times T \times F}.
\end{equation*}
\end{footnotesize}
Finally, $\mathbf{\ddot{U}^{(\text{c})}}$ is added to the input tensor via a residual connection to produce the output tensor: $\mathbf{\ddot{H}_X^{(\text{c})}}=\mathbf{\dot{H}_x^{(\text{c})}}+\mathbf{\ddot{U}^{(\text{c})}}$.

\subsubsection{Full- \& sub-band modeling} 
In the full- \& sub-band modeling block, time-dimension and frequency-dimension attention are employed to guide the models to focus on position (time frames) and content (frequency channel) respectively~\cite{zhang2022time}. 
Noteworthy, the attention module in our work shares the same network architecture as in~\cite{wang2022tf} to reduce the number of parameters of the proposed \model{}.

More specifically, taking `Time-MHA' in the sub-band modeling as an example, the input tensor $\mathbf{\ddot{H}_X^{(\text{c})}}$ is fed into a Conv2D with kernel $1\times 1$ followed by PReLU and LN along the channel and time dimensions (denoted as ctLN), then reshape operation is applied to form $\mathbf{Q_{\ell,t}}\in \mathbb{R}^{F\times(T\times E)}$, $\mathbf{K_{\ell,t}}\in \mathbb{R}^{F\times(T\times E)}$, $\mathbf{V_{\ell,t}}\in \mathbb{R}^{F\times(T\times D/L)}$:
\vspace{-0.5em}
\begin{equation*}\
\vspace{-0.5em}
    \begin{split}
    & \mathbf{Q_{\ell, t}^{(\text{c})}} = \operatorname{ctLN}(\operatorname{PReLU}(\operatorname{Conv2D}(\mathbf{\ddot{H}_X^{(\text{c})}}, D, E))),\\
    & \mathbf{K_{\ell, t}^{(\text{c})}} = \operatorname{ctLN}(\operatorname{PReLU}(\operatorname{Conv2D}(\mathbf{\ddot{H}_X^{(\text{c})}}, D, E))),\\
    & \mathbf{V_{\ell, t}^{(\text{c})}} = \operatorname{ctLN}(\operatorname{PReLU}(\operatorname{Conv2D}(\mathbf{\ddot{H}_X^{(\text{c})}}, D, D/L))),
    \end{split}
\end{equation*}
where $E$ is an embedding dimension that can be manually designated, $L$ is the number of heads in ``MHA''. 
After that, attention output $\mathbf{A_{\ell,t}}\in \mathbb{R}^{F\times(T\times D/L)}$ is computed as:
\vspace{-0.5em}
\begin{equation*}
\vspace{-0.5em}
\mathbf{A_{\ell,t}}=\operatorname{softmax}\left(\frac{\mathbf{Q_{\ell,t}}
\mathbf{K_{\ell,t}^{\top}}}{\sqrt{T \times E}}\right) \mathbf{V_{\ell,t}}.
\end{equation*}
We then concatenate the attention of all heads along the second dimension and reshape it back to $D\times T \times F$. At last, $1\times 1$ Conv2D with fixed input and output channels $D$ followed by PReLU and ctLN is applied to aggregate cross-head information, add it to the input tensor $\mathbf{\ddot{H}_X^{(\text{c})}}$ via a residual connection to produce the output tensor $\mathbf{\dddot{H}_X^{(\text{c})}}$.

The full-band modeling block and `Freq-MHA' contained within it share almost the same architecture as that in sub-band modeling block. The difference is that the modeling is processed within each temporal unit along the frequency dimensions, we need to change ctLN to cfLN (LN along the channel and frequency dimensions) and the reshaped dimensions of $\mathbf{Q_{\ell, f}}$, $\mathbf{K_{\ell, f}}$, $\mathbf{V_{\ell, f}}$ are $T\times (F\times E)$, $T\times (F\times E)$ and $T\times (F\times D/L)$ respectively.
\vspace{-0.5em}
\subsubsection{Multi-task learning}
To ensure the proposed \model{} optimizes both discriminative speaker embedding and the target speech, a multi-task learning framework with two objectives is introduced. To be specific, the scale-invariant signal-to-noise ratio (SI-SDR)~\cite{le2019sdr} loss measuring the quality between the extracted and clean target speech and the cross-entropy (CE) loss used for speaker classification is combined to optimize the network:
\vspace{-1em}

\begin{equation*}
\vspace{-0.5em}
\mathcal{L}=\mathcal{L}_{\text{SI-SDR}}(\mathbf{\hat{s}}, \mathbf{s})+\gamma \sum_{c=1}^C\mathcal{L}_{\text{CE}}(\mathbf{{\hat{y}^{(\text{c})}}}, \mathbf{y^{(\text{c})}}),
\end{equation*}
where $\mathbf{\hat{s}}$ and $\mathbf{s}$ denote the estimated and ground truth target speech, $\mathbf{\hat{y}^{(\text{c})}}$ and $\mathbf{{y^{(\text{c})}}}$ are the estimated and ground truth target speaker label. 
$\gamma$ is a scaling factor and set to $0.1$ in this paper.
\vspace{-0.5em}
\begin{equation*}
\vspace{-0.5em}
\mathbf{\hat{y}^{(\text{c})}} = \operatorname{Linear}({\mathbf{E^{(\text{c})}}}) \in \mathbb{R}^{(1,N)},
\end{equation*}
where $N$ is the number of speakers in the training dataset.
\vspace{-0.5em}
\section{Experiments}
\vspace{-0.5em}
\label{sec:exp}
\subsection{Datasets}
Each utterance in the datasets is simulated to be emitted from a specific location within a confined space. Therefore, the datasets include two parts: room impulse responses (RIRs) and speech.
\vspace{-1em}
\paragraph*{RIRs generation}
We use the randomized image method (RIM)~\cite{de2015modeling} to generate RIRs\footnote{https://github.com/LCAV/pyroomacoustics}.
Room dimensions in the RIR dataset are randomly generated, ranging from $3 \times 4 \times 2.13$ meters to $7 \times 8 \times 3$ meters. RT60 is also randomly generated and ranges from $0.1$ to $0.5$ seconds.
In each room, one microphone position and five speaker positions are randomly generated, with each position being at least $0.5$ meters away from the walls and floor and no higher than $1.8$ meters for increased realism. 
To balance the number of near and far sources, two of the speakers are placed near the microphone while the other three are placed far away. 
Near and far sources are distinguished based on a fixed threshold of $1.5$ meters.

\vspace{-1em}
\begin{figure}[htbp]
\centering
\includegraphics[width=1\linewidth]{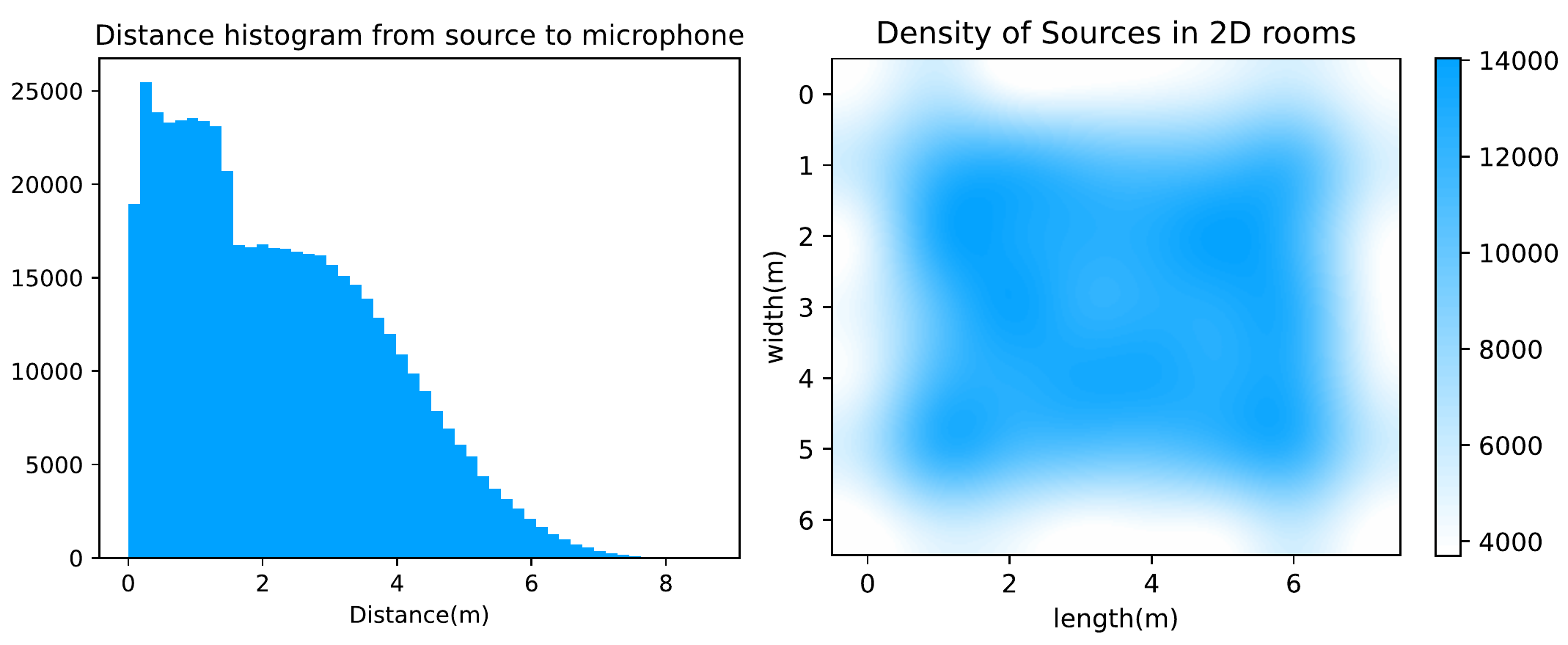}
\vspace{-2em}
\caption{Training data distributions for RIRs dataset. Distance distribution from microphone (left), spatial distribution (right)}
\label{fig3}
\vspace{-2em}
\end{figure}   

\paragraph*{Speech}
We use the small subset of LibriLight~\cite{kahn2020libri} containing about 577 hours of untranscribed speech from 489 speakers for training. 
Regarding validation and test datasets, we employ the ``dev-clean'' and ``test-clean'' subsets of Librispeech~\cite{panayotov2015librispeech}, each of which comprises 5.4 hours of speech from 40 speakers. 
The speech in the dataset is recorded at a sampling rate of 16kHz.
\paragraph*{Sample creation}
Applying randomization to the loudness of the speech is necessary. 
Specifically, the root mean square (RMS) energy of each speech signal is randomly set between $(-30, -20)$ dB before summing up all sources.
For the ablations in ~\Cref{sec: ablation study}, RMS of the speech beyond the threshold distance is randomization between $(-30, -10)$ dB to simulate more challenging scenarios, where speakers who are situated far away may potentially raise their voices in speech.
When discussing the n-Spkr dataset, the typical reference is to the presence of one speaker situated within the threshold distance, while there exist (n-1) speakers positioned beyond the threshold distance. 
Finally, a sample is obtained by convolving the RIR with the respective speech signal.
\vspace{-1em}
\subsection{Setup}
\vspace{-0.5em}
The number of the layers of extractor $C$ is set to $6$, while embedding dimensions of TF-units $D$ is $24$. Inside the `Time-MHA' and `Freq-MHA' blocks, embedding dimensions $E$ and the number of heads $L$ are both set to 4. For STFT, the window length is $16$ ms and hop length $8$ ms, a $256$-point discrete Fourier transform (DFT) is applied to extract $129$-dimensional complex STFT spectra at each frame.

Training runs with a batch size of 16 for at most 100 epochs using AdamW optimization~\cite{kingma2014adam} with a starting learning rate of 0.001, which is then gradually decreased using cosine annealing. Training stops when no improvement has been seen for more than 5 epochs.
\vspace{-1em}
\begin{table}[!htbp]
\centering
\caption{\model{} shows consistent improvement over LSTM and U-Net implementations on LibriSpeech dataset.}
\resizebox{20em}{!}{
\begin{tabular}{llrrr}
\toprule
Dataset  & Network & SI-SDR & SI-SDRi & PESQ \\
\midrule
\midrule
\multirow{4}{*}{2-Spkr}
& Mixture  &  5.02 & - &  1.541 \\  
& LSTM  &  10.02 & 5.00 & 1.917 \\
 & U-Net & 11.13 & 6.11& 2.088  \\
 & \model{} & \textbf{13.77} & \textbf{8.75} & \textbf{2.520}  \\
    \midrule 
\multirow{4}{*}{3-Spkr}
& Mixture  &  0.34 & -&  1.280 \\
& LSTM & 3.99 &  3.65 & 1.463  \\
 & U-Net & 5.21 & 4.87 & 1.570 \\
 & \model{} & \textbf{7.16} & \textbf{6.82} & \textbf{1.759}  \\
 \midrule 
 \multirow{4}{*}{4-Spkr} 
 & Mixture  &  -2.48 & - & 1.196 \\
 & LSTM & 0.29 & 2.77 & 1.305  \\
 & U-Net & 1.63& 4.11 & 1.380  \\
 & \model{} & \textbf{2.86} & \textbf{5.34} & \textbf{1.486}  \\
 \bottomrule
\end{tabular}}
\label{tab0}
\end{table}
\vspace{-2em}
\subsection{Comparison with other baseline models}
\vspace{-0.5em}
\label{sec:comparison_with_baseline}
We first compare the objective performance of \model{} with the baseline speech separation model, where LSTM follows the configuration from~\cite{patterson22_interspeech}. 
Also, we use a standard U-Net~\cite{ronneberger2015u} model as another baseline model, which is a lightweight 10-layer model with five encoder and five decoder layers, the number of filters for a layer for the encoder/decoder is $16, 32, 64, 128, 256$. 
Note that the training and validation set only contains two speakers (2-Spkr) while testing involves multiple speakers. 
Use SI-SDR as the loss function for the baseline model, as shown in~\Cref{tab0}, \model{} outperforms other baselines on all of the 2-, 3-, and 4-Speaker datasets.

\begin{table}[!htbp]
    \centering
    \caption{Ablation study, ``SE", ``T-Att'' and ``F-Att'' refer to speaker encoder, the sub-band modeling block consisting of `Time-MHA` and the full-band modeling block consisting of `Freq-MHA' respectively. Results in bold denote the best-achieved performance.} 
    \label{tab:ablation_1}
    \vspace{-0.5em}
    \resizebox{\linewidth}{!}{
    \begin{tabular}{lrrrrrrr}
    \toprule
        \multirow{2}{*}{Network}  &  \multicolumn{3}{c}{2-Spkr} & \multicolumn{3}{c}{3-Spkr}  \\
    \cmidrule(lr){2-4} \cmidrule(lr){5-7} 
    &  SI-SDR & SI-SDRi & PESQ &   SI-SDR & SI-SDRi & PESQ   \\
    \midrule
    \midrule
    Mixture & -0.04& - & 1.218 &-3.40  & -& 1.104\\
    \model{}    & \textbf{10.84}&  \textbf{10.88} & \textbf{2.103} &\textbf{3.07} & \textbf{6.47}  & \textbf{1.332} \\
    - w/o SE    & 10.28&  10.32 & 1.927 & 0.04 & 3.44 & 1.182 \\
    - w/o T-Att & 9.78&  9.82  & 1.930 & 1.88& 5.28 & 1.287 \\
    - w/o F-Att & 9.97&  10.01 & 2.088 & 2.03 & 5.43 & 1.308 \\
    \bottomrule
    \end{tabular}}
    \vspace{-2em}
\end{table}
\vspace{-1em}
\subsection{Ablation studies}
\vspace{-0.5em}
\label{sec: ablation study}
To determine the effectiveness of the improved method proposed in this paper, we study variants of \model{}. 
In this section, the training and validation set both contain two speakers (2-Spkr dataset) with and without an intruded speaker within the threshold distance. 
The duration of the intrusive speech is between 1 and 3 seconds, while the intruder appears at the end of the 5-second audio mixture. 
\Cref{tab:ablation_1} shows the performance of these variants, which demonstrates that the absence of any module results in a decrease in the overall performance of \model{}. 
It is worth noting that the variant without a speaker encoder shows a relatively significant decrease in performance on the 3-Spkr dataset, which suggests that the speaker encoder plays a significant role in multi-speaker scenarios.

We carried out further ablation experiments on the cross-dataset to better understand the impact of the speaker encoder. 
Three intricate scenarios are designed, the first involved interfering speakers within the extraction threshold distance, the second has speakers in a room with fainter reverberation (RT60 $\subseteq [0.1, 0.2]$s), and the third blends the characteristics of the former two scenes, namely the intrusion of the speaker and fainter reverberation.
Results in~\Cref{tab:ablation_2} demonstrate that the introduction of a speaker encoder can effectively mitigate such interference in the presence of interfering speakers within the threshold distance. Moreover, the \model{}'s performance remains strong even in rooms with shorter RT60. 
\vspace{-0.5em}
\begin{table}[!htbp]
    \centering
    \caption{Ablation study of speaker encoder in various complex scenarios. ``SE'' denotes speaker encoder, ``Faint'' RIRs mean that RT60 is shorter, ``Intruded'' speech means there are interfering speakers within the extraction threshold distance.
    } 
    \vspace{-1em}
    \resizebox{22em}{!}{
\begin{tabular}{llcrrr}
\toprule
\multicolumn{2}{c}{Dataset}  &  \multirow{2}{*}{Use SE?} &   \multirow{2}{*}{SI-SDR } &  \multirow{2}{*}{SI-SDRi} &  \multirow{2}{*}{PESQ} \\
    \cmidrule(lr){1-2}
RIRs & Speech  \\
\midrule
\midrule
\multirow{2}{*}{Normal} & \multirow{2}{*}{Unintruded}
&  \CheckmarkBold  & \textbf{10.84} &  \textbf{10.88} & \textbf{2.103}  \\
 & & \XSolidBrush  & 10.28 &   10.32 & 1.927\\
 
 \midrule
\multirow{2}{*}{Normal} & \multirow{2}{*}{Intruded}
&  \CheckmarkBold  &\textbf{8.40} &  \textbf{11.84} & \textbf{1.628}  \\
 & & \XSolidBrush  & 0.09 &   3.53 & 1.323\\

\midrule
\multirow{2}{*}{Faint} & \multirow{2}{*}{Unintruded}
&  \CheckmarkBold & \textbf{13.78} &  \textbf{7.79}  & \textbf{2.592} \\
 & & \XSolidBrush & 13.38 &  7.39  & 2.275\\

 \midrule
\multirow{2}{*}{Faint} & \multirow{2}{*}{Intruded}
& \CheckmarkBold & \textbf{7.16} &   \textbf{10.02}   & \textbf{1.900} \\
 & & \XSolidBrush & -1.20 &  1.39 & 1.428\\
 \bottomrule
\end{tabular}}
\label{tab:ablation_2}
\end{table}
\vspace{-1.5em}
\section{Conclusions}
This work\footnote{Demo: https://thuhcsi.github.io/interspeech2023-NS-Extractor/} introduced \model{}, a joint speaker and distance separation model for monaural TSE.
\model{} is a carefully designed model, based on the previously introduced TF-GridNet, optimized towards usage within different meeting scenarios.
Experimental results on several datasets that closely resemble real-life scenarios such as faint reverberation and unexpected intrusive speech demonstrate the efficacy of \model{} in complex scenarios.

\vspace{1em}

\textbf{Acknowledgements:}
This work is supported by National Natural Science Foundation of China (62076144), the Major Key Project of PCL (PCL2021A06, PCL2022D01) and Shenzhen Science and Technology Program (WDZC20220816140515001).

\clearpage

\bibliographystyle{IEEEtran}
\bibliography{mybib}
\end{document}